\documentclass[conference]{IEEEtran}
\IEEEoverridecommandlockouts
\usepackage{cite}
\usepackage{amsmath,amssymb,amsfonts}
\usepackage{algorithmic}
\usepackage{graphicx}
\usepackage[caption=false, font=footnotesize]{subfig}
\graphicspath{{figures/}}
\DeclareGraphicsExtensions{.pdf,.jpeg,.png}
\usepackage{textcomp}
\usepackage{xcolor}
\usepackage{url}
\usepackage{pifont}
\usepackage{enumitem}
\usepackage{hyperref}
\usepackage{physics}

\usepackage{tikz}
\usepackage{qcircuit}

\usepackage[inkscapepath=figures,inkscapearea=page]{svg}
\usepackage{tikz}
\usepackage{tikzscale}
\usetikzlibrary{arrows.meta,calc,decorations.markings,math,arrows.meta,arrows,intersections}
\usetikzlibrary{colorbrewer}
\usetikzlibrary{shapes.geometric}
\usetikzlibrary{automata, positioning,shapes.callouts}

\usepackage{tikzit}

\tikzstyle{none-small}=[fill=none, draw=none, shape=circle, tikzit category=misc, tikzit shape=circle, tikzit fill=none, font={\footnotesize}]
\tikzstyle{none-small-gray}=[fill=none, draw=none, shape=circle, text=gray, tikzit category=misc, tikzit shape=circle, tikzit fill=none, font={\footnotesize}]
\tikzstyle{gate}=[shape=rectangle, text height=1ex, text depth=0.25ex, yshift=0.5mm, fill=white, draw=black, minimum height=3mm, yshift=-0.5mm, minimum width=3mm, font={\footnotesize}, tikzit category=circuit]
\tikzstyle{meter}=[shape=rectangle, text height=1ex, text depth=0.25ex, yshift=0.5mm, fill=white, draw=black, minimum height=3mm, yshift=-0.5mm, minimum width=3mm, font={\footnotesize}, tikzit category=circuit, text width=4.5mm, label={{[shift={(0,-1.15)}]\metersymb}}]
\tikzstyle{big gate}=[shape=rectangle, text height=1.5ex, text depth=0.25ex, yshift=0.5mm, fill=white, draw=black, minimum height=10mm, yshift=-0.5mm, minimum width=5mm, font={\normalsize}, tikzit category=circuit]
\tikzstyle{long gate}=[shape=rectangle, text height=1ex, text depth=0.25ex, yshift=0.5mm, fill=white, draw=black, minimum height=3mm, yshift=-0.5mm, minimum width=5mm, font={\footnotesize}, tikzit category=circuit]
\tikzstyle{Z dot}=[inner sep=0mm, minimum size=2mm, shape=circle, draw=black, fill={rgb,255: red,221; green,255; blue,221}, tikzit category=zx]
\tikzstyle{Z phase dot}=[minimum size=5mm, font={\footnotesize\boldmath}, shape=rectangle, rounded corners=2mm, inner sep=1mm, outer sep=-2mm, scale=0.8, tikzit shape=circle, draw=black, fill={rgb,255: red,221; green,255; blue,221}, tikzit draw=blue, tikzit category=zx]
\tikzstyle{X dot}=[Z dot, shape=circle, draw=black, fill={rgb,255: red,255; green,136; blue,136}, tikzit category=zx]
\tikzstyle{X phase dot}=[Z phase dot, tikzit shape=circle, tikzit draw=blue, fill={rgb,255: red,255; green,136; blue,136}, font={\footnotesize\boldmath}, tikzit category=zx]
\tikzstyle{hadamard}=[fill=yellow, draw=black, shape=rectangle, inner sep=0.6mm, minimum height=1.5mm, minimum width=1.5mm, tikzit category=zx]
\tikzstyle{paulibox}=[fill={rgb,255: red,221; green,221; blue,255}, draw=black, shape=rectangle, inner sep=0.6mm, minimum height=5mm, minimum width=5mm, font={\footnotesize}, text height=1.5ex, text depth=0.25ex, tikzit category=zx]
\tikzstyle{vertex}=[inner sep=0mm, minimum size=1mm, shape=circle, draw=black, fill=black, tikzit category=misc]
\tikzstyle{vertex set}=[inner sep=0mm, minimum size=1mm, shape=circle, draw=black, fill=white, font={\footnotesize\boldmath}, tikzit category=misc]
\tikzstyle{small black dot}=[fill=black, draw=black, shape=circle, inner sep=0pt, minimum width=1.2mm, tikzit category=circuit]
\tikzstyle{cnot ctrl}=[fill=black, draw=black, shape=circle, inner sep=0pt, minimum width=1.2mm, tikzit category=circuit]
\tikzstyle{cnot targ}=[fill=white, draw=white, shape=circle, tikzit category=circuit, label={center:$\oplus$}, inner sep=0pt, minimum width=2.1mm, tikzit fill={rgb,255: red,102; green,204; blue,255}, tikzit draw=black]
\tikzstyle{ket}=[fill=white, draw=black, shape=regular polygon, regular polygon sides=3, regular polygon rotate=-30, scale=0.7, inner sep=1pt, tikzit category=circuit, tikzit shape=rectangle, tikzit fill=green]
\tikzstyle{bra}=[fill=white, draw=black, shape=regular polygon, regular polygon sides=3, regular polygon rotate=30, scale=0.7, inner sep=1pt, tikzit category=circuit, tikzit shape=rectangle, tikzit fill=red]
\tikzstyle{scalar}=[shape=rectangle, text height=1.5ex, text depth=0.25ex, yshift=0.5mm, fill=white, draw=black, minimum height=5mm, yshift=-0.5mm, minimum width=5mm, font={\normalsize}]
\tikzstyle{clabel}=[fill=white, draw=none, shape=rectangle, tikzit fill={rgb,255: red,56; green,255; blue,242}, font={\footnotesize}, inner sep=1pt, tikzit category=labels]
\tikzstyle{empty diagram}=[draw={gray!40!white}, dashed, shape=rectangle, minimum width=1cm, minimum height=1cm, tikzit category=misc]
\tikzstyle{cluster small}=[fill=none, thick, draw={rgb,255: red,0; green,128; blue,128}, shape=circle, tikzit category=misc, tikzit shape=circle, minimum size=1.5mm,  inner sep=0.3mm, tikzit fill=white, tikzit draw={rgb,255: red,0; green,128; blue,128}, font={\footnotesize}]
\tikzstyle{cluster}=[fill=none, thick, draw={rgb,255: red,0; green,128; blue,128}, shape=circle, tikzit category=misc, tikzit shape=circle, minimum size=3.5mm, inner sep=0pt, tikzit fill=white, tikzit draw={rgb,255: red,0; green,128; blue,128}, font={\footnotesize}]
\tikzstyle{cluster big}=[fill=none, thick, draw={rgb,255: red,0; green,128; blue,128}, shape=circle, tikzit category=misc, tikzit shape=circle, minimum size=4.5mm, text width=2mm, inner sep=0pt, tikzit fill=white, tikzit draw={rgb,255: red,0; green,128; blue,128}, font={\footnotesize}]

\tikzstyle{hadamard edge}=[-, dashed, dash pattern=on 2pt off 0.5pt, thick, draw={rgb,255: red,68; green,136; blue,255}]
\tikzstyle{box edge}=[-, dashed, dash pattern=on 2pt off 0.5pt, thick, draw={rgb,255: red,203; green,192; blue,225}]
\tikzstyle{brace edge}=[-, tikzit draw=blue, decorate, decoration={brace,amplitude=1mm,raise=-1mm}]
\tikzstyle{diredge}=[->]
\tikzstyle{double edge}=[-, double, shorten <=-1mm, shorten >=-1mm, double distance=2pt]
\tikzstyle{gray edge}=[-, {gray!70!white}, thick]
\tikzstyle{pointer edge}=[->, very thick, gray]
\tikzstyle{boldedge}=[-, line width=1.2pt, shorten <=-0.17mm, shorten >=-0.17mm]
\tikzstyle{boldedge red}=[-, line width=1.4pt, shorten <=-0.17mm, shorten >=-0.17mm, draw=red, tikzit draw=red]

\def\BibTeX{{\rm B\kern-.05em{\sc i\kern-.025em b}\kern-.08em
		T\kern-.1667em\lower.7ex\hbox{E}\kern-.125emX}}

\newcommand{\ghz}{G\!H\!Z}

\begin{document}
\bstctlcite{IEEEexample:BSTcontrol}
	
	\title{A compiler for universal photonic quantum computers
	}

\author{
	\IEEEauthorblockN{1\textsuperscript{st} Felix Zilk}
	\IEEEauthorblockA{\textit{Christian Doppler Laboratory}\\
		\textit{for Photonic Quantum Computer} \\
		\textit{Faculty of Physics} \\
		\textit{University of Vienna}\\
		Vienna, Austria\\
		felix.zilk@univie.ac.at}
	\\
	\IEEEauthorblockN{4\textsuperscript{th}Karl Fürlinger}
	\IEEEauthorblockA{\textit{MNM-Team} \\
		\textit{Ludwig-Maximilians-}\\
		\textit{Universität (LMU)}\\
		Munich, Germany \\
		fuerlinger@nm.ifi.lmu.de}
    \and
	\IEEEauthorblockN{2\textsuperscript{nd} Korbinian Staudacher}
	\IEEEauthorblockA{\textit{MNM-Team} \\
		\textit{Ludwig-Maximilians-}\\
		\textit{Universität (LMU)}\\
		Munich, Germany \\
		staudacher@nm.ifi.lmu.de}
	\\\\
	\IEEEauthorblockN{5\textsuperscript{th} Dieter Kranzlmüller}
	\IEEEauthorblockA{\textit{MNM-Team} \\
	\textit{Leibniz Supercomputing}\\
	\textit{Centre (LRZ)}\\
	Garching, Germany\\
	dieter.kranzlmueller@lrz.de}
    \and
	\IEEEauthorblockN{3\textsuperscript{rd} Tobias Guggemos}
	\IEEEauthorblockA{\textit{Christian Doppler Laboratory}\\
		\textit{for Photonic Quantum Computer} \\
		\textit{Faculty of Physics} \\
		\textit{University of Vienna}\\
		Vienna, Austria\\
		tobias.guggemos@univie.ac.at}
	\\
	\IEEEauthorblockN{6\textsuperscript{th} Philip Walther}
	\IEEEauthorblockA{\textit{Christian Doppler Laboratory}\\
	\textit{for Photonic Quantum Computer} \\
	\textit{Faculty of Physics} \\
	\textit{University of Vienna}\\
	Vienna, Austria\\
	philip.walther@univie.ac.at}
	}	

\maketitle

\makeatother   


\begin{abstract}
	Photons are a natural resource in quantum information, and the last decade showed significant progress in high-quality single photon generation and detection. 
	Furthermore, photonic qubits are easy to manipulate and do not require particularly strongly sealed environments, making them an appealing platform for quantum computing. 
	With the \textit{one-way model}, the vision of a universal and large-scale quantum computer based on photonics becomes feasible.
	In one-way computing, the input state is not an initial product state {\boldmath$ \ket{0}^{\otimes n} $}, but a so-called cluster state.
	A series of measurements on the cluster state's individual qubits and their temporal order, together with a  feed-forward procedure, determine the quantum circuit to be executed.
	We propose a pipeline to convert a \textit{QASM} circuit into a graph representation named \textit{measurement-graph} (\textit{m-graph}), that can be directly translated to hardware instructions on an optical one-way quantum computer.
	In addition, we optimize the graph using \textit{ZX-Calculus} before evaluating the execution on an experimental discrete variable photonic platform.
\end{abstract}

\begin{IEEEkeywords}
	Quantum Computing, Photonic QC, Measurement Based QC, One-way QC, ZX-Calculus
\end{IEEEkeywords}


\begin{figure*}[t]
	\centering
	\begin{tikzpicture}
	\begin{pgfonlayer}{nodelayer}
		\node [style=cluster] (0) at (4.75, 5.25) {1};
		\node [style=cluster] (1) at (6, 5.25) {2};
		\node [style=cluster] (2) at (7.25, 5.25) {3};
		\node [style=cluster] (3) at (8.5, 5.25) {4};
		\node [style=cluster] (4) at (13.75, 5.25) {1};
		\node [style=cluster] (5) at (15, 5.25) {2};
		\node [style=cluster] (6) at (15, 4) {3};
		\node [style=cluster] (7) at (13.75, 4) {4};
		\node [style=cluster] (8) at (10.75, 5.25) {1};
		\node [style=cluster] (9) at (12, 5.25) {2};
		\node [style=cluster] (10) at (12, 4) {3};
		\node [style=cluster] (11) at (10.75, 4) {4};
		\node [style=cluster] (24) at (16.75, 5.25) {1};
		\node [style=cluster] (25) at (18, 5.25) {2};
		\node [style=cluster] (26) at (16.75, 4) {4};
		\node [style=cluster] (27) at (18, 4) {3};
		\node [style=cluster] (28) at (19.75, 5.25) {1};
		\node [style=cluster] (29) at (21, 5.25) {2};
		\node [style=cluster] (30) at (19.75, 4) {4};
		\node [style=cluster] (31) at (21, 4) {3};
		\node [style=cluster] (32) at (22.75, 5.25) {1};
		\node [style=cluster] (33) at (24, 5.25) {2};
		\node [style=cluster] (34) at (24, 4) {3};
		\node [style=cluster] (35) at (22.75, 4) {4};
		\node [style=cluster small] (40) at (5.5, 2) {1};
		\node [style=cluster small] (41) at (6.75, 0.75) {4};
		\node [style=cluster small] (42) at (6.75, 2) {2};
		\node [style=cluster small] (43) at (8, 2) {3};
		\node [style=cluster small] (44) at (10.75, 2) {1};
		\node [style=cluster small] (45) at (12, 2) {2};
		\node [style=cluster small] (46) at (12, 0.75) {3};
		\node [style=cluster small] (47) at (10.75, 0.75) {4};
		\node [style=none-small] (48) at (6.75, 6) {``Linear''};
		\node [style=none-small] (49) at (6.75, 0) {$H_{1,3,4}\ket{GHZ}$};
		\node [style=none-small] (50) at (11.25, 6) {``Box''};
		\node [style=none-small] (51) at (6.75, 2.75) {``GHZ''};
		\node [style=none] (52) at (0.25, 3.25) {};
		\node [style=none] (53) at (24.75, 3.25) {};
		\node [style=none-small] (54) at (6.75, 4.25) {$H_{1,4}\ket{\psi}$};
		\node [style=none-small] (55) at (9.75, 4.75) {$\overset{l.c.}{=}$};
		\node [style=none-small] (56) at (13, 4.75) {$\overset{l.c.}{=}$};
		\node [style=none-small] (57) at (16, 4.75) {$\overset{l.c.}{=}$};
		\node [style=none-small] (58) at (19, 4.75) {$\overset{l.c.}{=}$};
		\node [style=none-small] (59) at (22, 4.75) {$\overset{l.c.}{=}$};
		\node [style=none-small] (60) at (9.5, 1.5) {$\overset{l.c.}{=}$};
		\node [style=none-small] (61) at (1.75, 4.75) {\textbf{Linear}};
		\node [style=none-small] (62) at (1.75, 1.25) {\textbf{GHZ}};
		\node [style=none-small] (64) at (1.75, 4) {\textbf{graph states}};
		\node [style=none-small] (65) at (1.75, 0.5) {\textbf{graph states}};
	\end{pgfonlayer}
	\begin{pgfonlayer}{edgelayer}
		\draw [style=boldedge] (0) to (1);
		\draw [style=boldedge] (1) to (2);
		\draw [style=boldedge] (2) to (3);
		\draw [style=boldedge] (4) to (5);
		\draw [style=boldedge] (5) to (6);
		\draw [style=boldedge] (6) to (7);
		\draw [style=boldedge] (8) to (9);
		\draw [style=boldedge] (9) to (10);
		\draw [style=boldedge] (10) to (11);
		\draw [style=boldedge] (8) to (11);
		\draw [style=boldedge] (5) to (7);
		\draw [style=boldedge] (24) to (25);
		\draw [style=boldedge, in=45, out=-135] (25) to (26);
		\draw [style=boldedge] (26) to (27);
		\draw [style=boldedge] (25) to (27);
		\draw [style=boldedge, in=135, out=-45] (24) to (27);
		\draw [style=boldedge] (28) to (29);
		\draw [style=boldedge, in=45, out=-135] (29) to (30);
		\draw [style=boldedge] (30) to (31);
		\draw [style=boldedge] (28) to (31);
		\draw [style=boldedge] (32) to (33);
		\draw [style=boldedge] (33) to (34);
		\draw [style=boldedge] (34) to (35);
		\draw [style=boldedge] (33) to (35);
		\draw [style=boldedge] (32) to (35);
		\draw [style=boldedge] (41) to (42);
		\draw [style=boldedge] (42) to (43);
		\draw [style=boldedge] (40) to (42);
		\draw [style=boldedge] (44) to (45);
		\draw [style=boldedge] (45) to (46);
		\draw [style=boldedge] (46) to (47);
		\draw [style=boldedge] (45) to (47);
		\draw [style=boldedge] (44) to (47);
		\draw [style=boldedge] (44) to (46);
		\draw [style=box edge] (52.center) to (53.center);
	\end{pgfonlayer}
\end{tikzpicture}
	\vspace{-1cm}
	\caption{Generation of the 4-qubit \textit{Linear} and \textit{GHZ} graph state from $ \ket{\psi} = \frac{1}{2} (\ket{0000} + \ket{0011} + \ket{1100} - \ket{1111}) $  and \mbox{$ \ket{G\!H\!Z} = \frac{1}{\sqrt{2}} (\ket{0000} + \ket{1111}) $}. The graph states in each are local complementary (l.c.) with each-other, and can be use as input states for the one-way-model.}
	\label{fig:cluster-states}
\end{figure*}
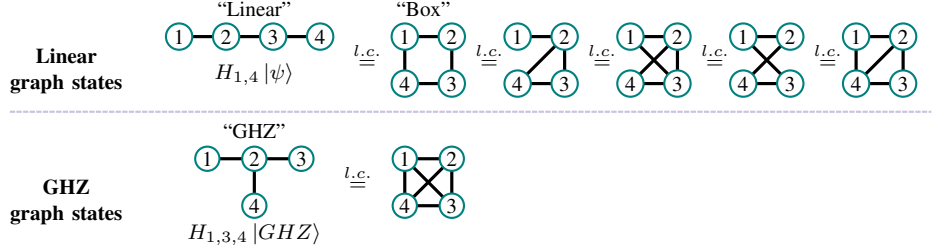

\section{Introduction}%
\label{sec:intr}
Photons are a natural candidate for quantum computing, yet such systems are not very prevalent in Cloud or High-Performance Computing (HPC) platforms.
However, photonic systems should be considered a valid competitor to other platforms; recent findings show setups with up to 14 entangled photons~\cite{Thomas.2022}.

HPC is in an era of specialization, where an increasing number of accelerator devices are integrated in general-purpose computing machines~\cite{shalf2019hpc}. 
Quantum computers represent an especially powerful type of accelerator, promising speed-ups for unstructured search and combinatorial optimization problems~\cite{cerezo2021variational,nielsenchuang,Nisq-algos}. 
As quantum technology matures, it is important to enable integration of quantum processing units (QPUs) in the HPC ecosystem and to support heterogeneous programming that integrates classical and quantum aspects, for example by means of offloading~\cite{chall+opp, mccaskey2020xacc, sivarajah2020t, oqasm, leymannqucloud, Karalekas_2020}. 
Here, quantum algorithms are usually expressed as offloaded computational kernels in a domain-specific language for the quantum circuit model (e.g., \textit{QASM})~\cite{LaRose2019, sivarajah2020t}. 

Although photons are an attractive platform for QPUs, a direct translation of the quantum circuit model into photonic components is impractical.
Photonic two-qubit gates are intrinsically probabilistic; hence, an increasing number of gates comes with an exponential decrease in the circuit's success probability.
That is why measurement-based schemes~\cite{Gao2011, Briegel2009ProgArt, Barz.2015} are an appealing alternative, in particular the \textit{one-way model} of quantum computing~\cite{Barz.2015, conciserev, Walther-oneway, Prevedel-feedforw}.
Here, computation is carried out solely by single-qubit measurements on highly-entangled multipartite states -- so-called \textit{cluster states}~\cite{raussendorf2001one}. 
The model is equivalent to the circuit model~\cite{Raussendorf2005}, but efficient methods for translation are still rare~\cite{PhotonicOneWayCompilerStBarb, RohdeGraphStateCompiler}.

\subsubsection*{Contribution}
This paper describes our efforts to develop a compiler for \textit{QASM} kernels that targets discrete variable photonic platforms.
We propose a pipeline to translate from \textit{QASM} to a graph representation, named \textit{measurement-graph}, or \textit{m-graph}.
The \textit{m-graph} is optimized with ZX-Calculus~\cite{coecke2018picturing} and mapped to hardware instructions for a photonic one-way processor.
This marks a first step towards accessing photonic QPUs with HPC systems.


\section{Background}
\label{sec:background}
The quantum circuit model performs computations by sequentially applying unitary gates to qubits in a quantum register.
Typically, one initializes a register of $ n $ qubits in the product state $ \ket{0}^{\otimes n}$~\cite{nielsenchuang}.

In contrast to the circuit model, the paradigm of quantum annealing~\cite{PerspectiveAnnealingHauke_2020} bases on the unitary evolution of the underlying system Hamiltonian. 

\subsection{The measurement-based one-way model}
The one-way model, however, bases entirely on adaptive single-qubit measurements that drive the computation~\cite{Barz.2015, Briegel2009ProgArt, Wei2021MeasurementBasedQC}. 
Here, the initial state of individual qubits is the $\ket{+}$ state, and they are pairwise coupled via $CZ$ operations to form a graph state, which serves as the computational resource. 
Then, single-qubit measurements on this resource state achieve universal quantum computation.

A \textit{cluster state} is a type of graph state in which the underlying graph structure has the form of a two-dimensional orthogonal grid.
\textit{Graph states} are highly-entangled multipartite states, and we represent them mathematically as a graph $G(V,E)$, with vertices $V$ representing physical qubits and edges $E$ indicating entanglement between qubits. 
An arbitrary graph state $ \ket{G} $ of $ V $ qubits and $ E $ edges is~\cite{Russo_2019}
\begin{equation}
 \ket{G} = \left(\prod_{(a,b) \in E} CZ_{a,b} \right) \bigotimes_{v\in V} \ket{+}_v
\end{equation}
\autoref{fig:cluster-states} shows a collection of 4-qubit cluster states, arranged as a two-dimensional lattice.
The states are \textit{locally complementary} (l.c.) if one can be transformed into the other with single-qubit transformations and \textit{SWAP} operations only.

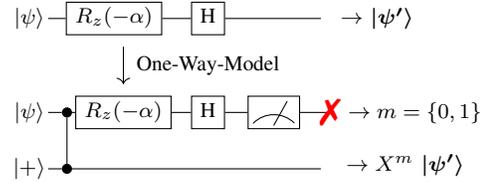
\begin{figure}[t]
    \centering
	\begin{tikzpicture}
	\begin{pgfonlayer}{nodelayer}
		\node [style=none-small] (1) at (-3.75, 1.5) {$\ket{\psi}$};
		\node [style=gate] (4) at (-1.5, 1.5) {$R_z(-\alpha)$};
		\node [style=gate] (5) at (1, 1.5) {H};
		\node [style=none] (6) at (4, 1.5) {};
		\node [style=none] (9) at (-3.25, 1.5) {};
		\node [style=none-small] (10) at (-3.75, -1) {$\ket{\psi}$};
		\node [style=gate] (11) at (-1.25, -1) {$R_z(-\alpha)$};
		\node [style=gate] (12) at (1, -1) {H};
		\node [style=none-small] (15) at (6.25, -2.4) {$\rightarrow X^{m}$ {\boldmath $\ket{\psi'}$}};
		\node [style=none] (16) at (-3.25, -1) {};
		\node [style=meter] (17) at (2.75, -1) {};
		\node [style=cnot ctrl] (18) at (-2.75, -1) {};
		\node [style=cnot ctrl] (19) at (-2.75, -2.5) {};
		\node [style=none-small] (20) at (-3.75, -2.5) {$\ket{+}$};
		\node [style=none] (21) at (-3.25, -2.5) {};
		\node [style=none] (22) at (4, -2.5) {};
		\node [style=none] (24) at (4.25, -1) {\color{red}\Large\ding{55}};
		\node [style=none] (25) at (4.25, -1) {};
		\node [style=none] (26) at (-1.25, 0.75) {};
		\node [style=none] (27) at (-1.25, -0.25) {};
		\node [style=none-small] (29) at (1, 0.25) {One-Way-Model};
		\node [style=none-small] (30) at (5.5, 1.5) {$\rightarrow$ {\boldmath $\ket{\psi'}$}};
		\node [style=none-small] (31) at (6.5, -1) {$\rightarrow m=\{0,1\}$};
	\end{pgfonlayer}
	\begin{pgfonlayer}{edgelayer}
		\draw (9.center) to (4);
		\draw (4) to (5);
		\draw (5) to (6.center);
		\draw (11) to (12);
		\draw (12) to (17);
		\draw (18) to (19);
		\draw (16.center) to (18);
		\draw (18) to (11);
		\draw (21.center) to (19);
		\draw (19) to (22.center);
		\draw [in=180, out=0] (17) to (25.center);
		\draw [style=diredge] (26.center) to (27.center);
	\end{pgfonlayer}
\end{tikzpicture}
	\caption{Circuit representation of the conceptual difference between the circuit model with unitary transformation and the one-way model. 
		Upper circuit: shows the transformation of a qubit in state $ \ket{\psi} $ with the unitary $ H R_z(-\alpha) $ to state $ \ket{\psi'} $.
		Lower circuit: shows the same transformation with the one-way model, where the new state $ \ket{\psi'} $ is than teleported to the bottom qubit by measuring the upper one (up to a Pauli-$X$ correction, that is based on the measurements output $ m $).}
	\label{eq:one-way}\label{fig:one-way}
\end{figure}

To carry out computation, we subsequentially measure on connected physical qubits $ j $ in the \textit{equatorial basis} $\mathcal{B}_{j}(\alpha)=\{\ket{\alpha}_{j},\ket{-\alpha}_{j}\}$, where \mbox{$\ket{\pm\alpha}_{j}=1/\sqrt{2}(\ket{0}_{j}\pm e^{i\alpha}\ket{1}_{j})$}. 
Measurements of physical qubits in the basis $\mathcal{B}_{j}(\alpha)$ induce the rotation $ H R_z\left(-\alpha\right)\ket{\psi} $ on encoded logical qubits up to a Pauli-$X$ correction (cf. \autoref{fig:one-way}).
$CZ$ gates are inherently built into the computational resource state as links between two qubits. 

Thus, the \textit{native gate set} in the one-way model may be defined as \mbox{$\mathcal{G}=\{H R_z(-\alpha), CZ\}$}, which is indeed universal.

Unlike the unitary evolution in the gate-based model, the non-unitary action of measurement is irreversible. 
As each measurement's outcome is random, the desired result only occurs in some cases.
Hence, a \textit{feed-forward} protocol~\cite{Prevedel-feedforw} compensates undesired results by adapting future measurement bases according to earlier outcomes.

As an example, refer to the single-qubit computation in \autoref{eq:one-way}.
The result $ m $ of the upper wire measurement (red cross mark) influences the Pauli-$X$ correction on the lower output wire.
If the outcome is $m=0$, the algorithm works as expected; however, if $m = 1$, a Pauli error is introduced and corrected before the final measurement.
The cascaded execution of this procedure allows for the implementation of arbitrary single-qubit rotations.
In fact, feed-forward control makes one-way quantum computation deterministic.

Furthermore, any quantum circuit can be converted to a measurement pattern on a sufficiently large cluster state~\cite{Barz.2015, Briegel2009ProgArt, Wei2021MeasurementBasedQC}.

\begin{table*}[t]
	\caption{Some elementary quantum gates and their respective representation in a ZX-diagram.}
	\label{fig:overview-zx}
	\setlength\arraycolsep{1.5pt}
\resizebox{\textwidth}{!}{
	\begin{tabular}{c | c c c c c c }
		\textbf{Name} & \textbf{Z} & \textbf{Z-Phase} & \textbf{X} & \textbf{X-Phase} & \textbf{CNOT} & \textbf{CZ}\\[.5em]\hline
		\textbf{Matrix} & $ \begin{pmatrix}1 & 0 \\ 0 & -1\end{pmatrix} $ 
		& \quad$ \begin{pmatrix}1 & 0 \\ 0 & e^{i\alpha}\end{pmatrix} $\quad
		& $ \begin{pmatrix}0 & 1 \\ 1 & 0\end{pmatrix} $  
		& $\frac{1}{2} \begin{pmatrix}1 + e^{i\alpha} & 1 - e^{i\alpha} \\ 1 - e^{i\alpha} & 1 + e^{i\alpha}\end{pmatrix} $  
		& $ \begin{pmatrix} 1 & 0 & 0 & 0 \\ 0 & 1 & 0 & 0 \\ 0 & 0 & 0 & 1 \\ 0 & 0 & 1 & 0 \end{pmatrix} $
		& $ \begin{pmatrix} 1 & 0 & 0 & 0 \\ 0 & 1 & 0 & 0 \\ 0 & 0 & 1 & 0 \\ 0 & 0 & 0 & -1 \end{pmatrix} $ \\[1em]
		\textbf{Gate} 
		& $ \Qcircuit @C=.5em @R=.75em {& \gate{Z} & \qw} $ 
		& $ \Qcircuit @C=.5em @R=.75em {& \gate{R_z(\alpha)} & \qw} $
		& $ \Qcircuit @C=.5em @R=.75em {& \gate{X} & \qw} $
		& $ \Qcircuit @C=.5em @R=.75em {& \gate{R_x(\alpha)} & \qw} $
		& $ \Qcircuit @C=.5em @R=1em {& \ctrl{1} & \qw \\ & \targ & \qw} $
		& $ \Qcircuit @C=.5em @R=1em {& \ctrl{1} & \qw \\ & \control\qw & \qw} $ \\
		\textbf{Spider/ Wire}  & \begin{tikzpicture}
	\begin{pgfonlayer}{nodelayer}
		\node [style=none] (0) at (0, 1) {};
		\node [style=Z phase dot] (1) at (0, 0) {$\pi$};
		\node [style=none] (2) at (0, -1) {};
	\end{pgfonlayer}
	\begin{pgfonlayer}{edgelayer}
		\draw (0.center) to (1);
		\draw (1) to (2.center);
	\end{pgfonlayer}
\end{tikzpicture}  & \begin{tikzpicture}
	\begin{pgfonlayer}{nodelayer}
		\node [style=Z phase dot] (0) at (0, 0) {$\alpha$};
		\node [style=none] (1) at (0, 1) {};
		\node [style=none] (2) at (0, -1) {};
	\end{pgfonlayer}
	\begin{pgfonlayer}{edgelayer}
		\draw (1.center) to (0);
		\draw (0) to (2.center);
	\end{pgfonlayer}
\end{tikzpicture}  & \begin{tikzpicture}
	\begin{pgfonlayer}{nodelayer}
		\node [style=none] (0) at (0, 1) {};
		\node [style=X phase dot] (1) at (0, 0) {$\pi$};
		\node [style=none] (2) at (0, -1) {};
	\end{pgfonlayer}
	\begin{pgfonlayer}{edgelayer}
		\draw (0.center) to (1);
		\draw (1) to (2.center);
	\end{pgfonlayer}
\end{tikzpicture} & \begin{tikzpicture}
	\begin{pgfonlayer}{nodelayer}
		\node [style=X phase dot] (0) at (0, 0) {$\alpha$};
		\node [style=none] (1) at (0, 1) {};
		\node [style=none] (2) at (0, -1) {};
	\end{pgfonlayer}
	\begin{pgfonlayer}{edgelayer}
		\draw (1.center) to (0);
		\draw (0) to (2.center);
	\end{pgfonlayer}
\end{tikzpicture} &  \begin{tikzpicture}
	\begin{pgfonlayer}{nodelayer}
		\node [style=Z dot] (0) at (0, 0.5) {};
		\node [style=X dot] (1) at (0, -0.5) {};
		\node [style=none] (2) at (-1, -0.5) {};
		\node [style=none] (3) at (-1, 0.5) {};
		\node [style=none] (4) at (1, 0.5) {};
		\node [style=none] (5) at (1, -0.5) {};
	\end{pgfonlayer}
	\begin{pgfonlayer}{edgelayer}
		\draw (4.center) to (0);
		\draw (0) to (1);
		\draw (1) to (5.center);
		\draw (1) to (2.center);
		\draw (0) to (3.center);
	\end{pgfonlayer}
\end{tikzpicture}	& \begin{tikzpicture}
	\begin{pgfonlayer}{nodelayer}
		\node [style=Z dot] (0) at (0, 0.5) {};
		\node [style=Z dot] (1) at (0, -0.5) {};
		\node [style=none] (2) at (-1, -0.5) {};
		\node [style=none] (3) at (-1, 0.5) {};
		\node [style=none] (4) at (1, 0.5) {};
		\node [style=none] (5) at (1, -0.5) {};
	\end{pgfonlayer}
	\begin{pgfonlayer}{edgelayer}
		\draw (4.center) to (0);
		\draw [style=hadamard edge] (0) to (1);
		\draw (1) to (5.center);
		\draw (1) to (2.center);
		\draw (0) to (3.center);
	\end{pgfonlayer}
\end{tikzpicture}		 
	\end{tabular}
}
\end{table*}

\subsection{Implementation with a photonic processor}
Photons are excellent candidates for building quantum computers; they are easy to generate and detect, robust against decoherence, and optical experiments can realize accurate single-qubit gates easily. 
However, deterministic interactions of two photons are experimentally impossible, and photonic two-qubit gates are of probabilistic nature~\cite{Barz.2015, conciserev}.

In the one-way model, these nondeterministic operations prepare the cluster state, just before any logical computation takes place~\cite{conciserev, Barz.2015}.

\textit{Post-selection} techniques ensure successful generation of cluster states, such that it ignores certain detection events from the results where the cluster state generation failed~\cite{Walther-oneway, Prevedel-feedforw}.

High-precision measurements in an arbitrary basis are achieved, for example, with phase retarders (wave plates) for qubits encoded in photon polarization.
After successfully generating the multi-photon cluster, the computation proceeds deterministically by sequential execution of such single-qubit measurements. 

Despite its limitations for implementing a pure circuital model, photonic qubits are well-suited to the one-way model.
Hence, the challenge to realizing universal photonic quantum computing lies in the efficient creation of sufficiently large cluster states.
Various such protocols have been proposed and implemented~\cite{sequcluster20, JWPanSetup, psiquantum-linopt-cluster, Thomas.2022}. 
In fact, smaller clusters can already implement interestingly large circuits such as the Grover search in \autoref{fig:grover}.

\begin{figure}[h]
	\centering
  	\small
	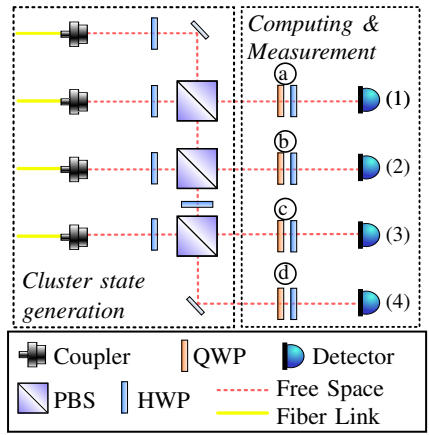
	\caption{A photonic quantum processing unit (QPU) to execute a \textit{graph state} of 4 nodes, as in \autoref{fig:cluster-states}.
		The input are four single-photons in four fiber links. Half-wave plates (HWPs) initialize their states to $\ket{D} \equiv \ket{+}$ before they are pairwise entangled by polarizing beam splitters (PBSs) to create the cluster state.
		The combination of a quarter-wave plate (QWP) and a HWP before the detectors allow measurements in arbitrary bases. }
	\label{fig:setup}
\end{figure}

\autoref{fig:setup} shows how a photonic QPU can run a quantum circuit on a cluster state created from four photons.
The input state generation step may be viewed independently from the computing and measurement steps that executes the necessary measurements for the post-selection and information processing.

\subsection{ZX-Calculus}
The ZX-Calculus is a diagrammatic language for quantum computation. 
A ZX-diagram consists of spiders and wires, which represent linear maps. 
Like the known quantum circuit notation, wires entering the diagram from the left are called input wires, and wires exiting to the right are called output wires. 

Spiders are distinguished into Z (green) and X (red) spiders and can have any number of input and output wires. 
In Dirac notation, their linear maps are: 
\begin{equation}
	\begin{tikzpicture}
	\begin{pgfonlayer}{nodelayer}
		\node [style=Z phase dot] (0) at (-2.75, 1) {$\alpha$};
		\node [style=none] (1) at (-1.75, 1.75) {};
		\node [style=none] (2) at (-1.75, 0.25) {};
		\node [style=none] (3) at (-3.75, 0.25) {};
		\node [style=none] (4) at (-3.75, 1.75) {};
		\node [style=none] (5) at (-3.9, 1.25) {$\vdots$};
		\node [style=none] (6) at (-1.6, 1.25) {$\vdots$};
		\node [style=X phase dot] (7) at (2.5, 1) {$\alpha$};
		\node [style=none] (8) at (3.5, 1.75) {};
		\node [style=none] (9) at (3.5, 0.25) {};
		\node [style=none] (10) at (1.5, 0.25) {};
		\node [style=none] (11) at (1.5, 1.75) {};
		\node [style=none] (12) at (1.6, 1.25) {$\vdots$};
		\node [style=none] (13) at (3.4, 1.25) {$\vdots$};
		\node [style=none-small] (18) at (-2.5, -0.5) {$:=\ket{0}^{\otimes m}\!\bra{0}^{\otimes n}$};
		\node [style=none-small] (23) at (-2.8, -1.15) {\footnotesize$+ e^{i\alpha}\ket{1}^{\otimes m}\!\bra{1}^{\otimes n}$};
		\node [style=none-small] (24) at (3, -0.5) {$:=\ket{+}^{\otimes m}\!\bra{+}^{\otimes n}$};
		\node [style=none-small] (25) at (2.7, -1.15) {$+ e^{i\alpha}\ket{-}^{\otimes m}\!\bra{-}^{\otimes n}$};
		\node [style=none-small] (26) at (-4.5, 1) {$m$};
		\node [style=none-small] (27) at (-1, 1) {$n$};
		\node [style=none-small] (28) at (1, 1) {$m$};
		\node [style=none-small] (29) at (4, 1) {$n$};
		\node [style=none] (31) at (1.5, 0.75) {};
		\node [style=none] (32) at (0, 1.75) {};
		\node [style=none] (33) at (0, -1.5) {};
	\end{pgfonlayer}
	\begin{pgfonlayer}{edgelayer}
		\draw [in=210, out=30] (0) to (1.center);
		\draw [in=150, out=-30] (0) to (2.center);
		\draw [in=30, out=-150] (0) to (3.center);
		\draw [in=330, out=150] (0) to (4.center);
		\draw [in=210, out=30] (7) to (8.center);
		\draw [in=150, out=-30] (7) to (9.center);
		\draw [in=30, out=-150] (7) to (10.center);
		\draw [in=330, out=150] (7) to (11.center);
		\draw [style=gray edge] (32.center) to (33.center);
	\end{pgfonlayer}
\end{tikzpicture}
	\vspace{-1.3cm}
\end{equation}
Apart from the normal wire corresponding to $\mathbb{I}_2$, it is convenient to introduce another wire type for representing the Hadamard matrix:
\begin{equation}
	\begin{tikzpicture}
	\begin{pgfonlayer}{nodelayer}
		\node [style=none] (0) at (-4, 1) {};
		\node [style=none] (1) at (-4, -1) {};
		\node [style=none] (2) at (-3, 0) {$:=$};
		\node [style=none] (3) at (-1, 0) {$\begin{pmatrix} 1 & 0 \\ 0 & 1 \end{pmatrix}$};
		\node [style=none] (4) at (2, 1) {};
		\node [style=none] (5) at (2, -1) {};
		\node [style=none] (6) at (3, 0) {$:=$};
		\node [style=none] (7) at (6, 0) {$\frac{1}{\sqrt{2}}\begin{pmatrix} 1 & 1 \\ 1 & -1 \end{pmatrix}$};
	\end{pgfonlayer}
	\begin{pgfonlayer}{edgelayer}
		\draw (0.center) to (1.center);
		\draw [style=hadamard edge] (4.center) to (5.center);
	\end{pgfonlayer}
\end{tikzpicture}
\end{equation}
Two diagrams can be composed horizontally by joining the output wires of one diagram to the input wires of the other, or vertically by placing them side by side. 
In this way, any quantum circuit can be represented as a ZX-diagram, moreover, most elementary quantum gates can be directly translated, as shown in \autoref{fig:overview-zx}.

\begin{figure*}[t]
	\centering
	\subfloat[\label{fig:gl-diagram}]{\begin{tikzpicture}[baseline={(0,0)}]
	\begin{pgfonlayer}{nodelayer}
		\node [style=none] (0) at (0, 3.5) {};
		\node [style=gate] (1) at (1.5, 3.5) {$R_x(\alpha)$};
		\node [style=gate] (2) at (3.75, 3.5) {$R_z(\beta)$};
		\node [style=gate] (3) at (6, 3.5) {$R_x(\gamma)$};
		\node [style=none] (4) at (7.5, 3.5) {};
		\node [style=none] (5) at (3.75, 2.5) {$=$};
		\node [style=none] (11) at (3.75, 0.75) {$=$};
		\node [style=none] (12) at (0, 1.75) {};
		\node [style=X phase dot] (13) at (1.5, 1.75) {$\alpha$};
		\node [style=Z phase dot] (14) at (3.75, 1.75) {$\beta$};
		\node [style=X phase dot] (15) at (6, 1.75) {$\gamma$};
		\node [style=none] (16) at (7.5, 1.75) {};
		\node [style=none] (17) at (0, 0) {};
		\node [style=Z phase dot] (18) at (1.5, 0) {$\alpha$};
		\node [style=Z phase dot] (19) at (3.75, 0) {$\beta$};
		\node [style=Z phase dot] (20) at (6, 0) {$\gamma$};
		\node [style=none] (21) at (7.5, 0) {};
	\end{pgfonlayer}
	\begin{pgfonlayer}{edgelayer}
		\draw (0.center) to (1);
		\draw (1) to (2);
		\draw (2) to (3);
		\draw [in=180, out=0] (3) to (4.center);
		\draw (12.center) to (13);
		\draw [in=180, out=0] (13) to (14);
		\draw (14) to (15);
		\draw [in=180, out=0] (15) to (16.center);
		\draw [style=hadamard edge] (17.center) to (18);
		\draw [style=hadamard edge] (18) to (19);
		\draw [style=hadamard edge] (19) to (20);
		\draw [style=hadamard edge, in=180, out=0] (20) to (21.center);
	\end{pgfonlayer}
\end{tikzpicture}}\hfill
	\subfloat[\label{fig:gl-rules-lc}]{\begin{tikzpicture}
	\begin{pgfonlayer}{nodelayer}
		\node [style=Z phase dot] (0) at (3, 1.75) {$\alpha_1$};
		\node [style=Z phase dot] (1) at (1, 1.75) {$\pm\frac{\pi}{2}$};
		\node [style=Z phase dot] (2) at (1, 0) {$\alpha_2$};
		\node [style=Z phase dot] (3) at (3, 0) {$\alpha_3$};
		\node [style=Z phase dot] (19) at (8, 1.75) {$\alpha_1\mp\frac{\pi}{2}$};
		\node [style=Z phase dot] (21) at (5.75, 0) {$\alpha_2\mp\frac{\pi}{2}$};
		\node [style=Z phase dot] (22) at (8, 0) {$\alpha_3\mp\frac{\pi}{2}$};
		\node [style=none] (23) at (1, 2.5) {$a$};
		\node [style=none] (24) at (4.5, 1.25) {$\overset{(lc)}{=}$};
		\node [style=none] (25) at (0.25, 2.75) {};
		\node [style=none] (26) at (0.25, 1.25) {};
		\node [style=none] (27) at (1.75, 2.75) {};
		\node [style=none] (28) at (1.75, 1.25) {};
	\end{pgfonlayer}
	\begin{pgfonlayer}{edgelayer}
		\draw [style=hadamard edge] (1) to (0);
		\draw [style=hadamard edge] (1) to (2);
		\draw [style=hadamard edge] (1) to (3);
		\draw [style=hadamard edge] (21) to (19);
		\draw [style=hadamard edge] (2) to (3);
		\draw [style=hadamard edge] (19) to (22);
		\draw [style=gray edge] (25.center) to (27.center);
		\draw [style=gray edge] (27.center) to (28.center);
		\draw [style=gray edge] (28.center) to (26.center);
		\draw [style=gray edge] (26.center) to (25.center);
	\end{pgfonlayer}
\end{tikzpicture}}\hfill
	\subfloat[\label{fig:gl-rules-p}]{\begin{tikzpicture}
	\begin{pgfonlayer}{nodelayer}
		\node [style=Z phase dot] (0) at (0, 0) {$\alpha_2$};
		\node [style=Z phase dot] (1) at (0, 1.75) {$\alpha_1$};
		\node [style=Z phase dot] (2) at (1.25, 1.75) {$j\pi$};
		\node [style=Z phase dot] (3) at (3.75, 1.75) {$k\pi$};
		\node [style=Z phase dot] (4) at (5, 1.75) {$\gamma_1$};
		\node [style=Z phase dot] (5) at (2.5, 0) {$\beta_1$};
		\node [style=Z phase dot] (6) at (5, 0) {$\gamma_2$};
		\node [style=none] (7) at (6.25, 1) {$\overset{(p)}{=}$};
		\node [style=Z phase dot] (8) at (7.5, 1.75) {$\alpha_1'$};
		\node [style=Z phase dot] (9) at (7.5, 0) {$\alpha_2'$};
		\node [style=Z phase dot] (12) at (10.5, 0) {$\gamma_2'$};
		\node [style=Z phase dot] (13) at (9, 2.75) {$\beta_1'$};
		\node [style=Z phase dot] (14) at (10.5, 1.75) {$\gamma_1'$};
		\node [style=none] (16) at (1.25, 2.5) {$u$};
		\node [style=none] (17) at (3.75, 2.5) {$v$};
		\node [style=none] (21) at (0.75, 2.75) {};
		\node [style=none] (22) at (0.75, 1.25) {};
		\node [style=none] (23) at (4.25, 2.75) {};
		\node [style=none] (24) at (4.25, 1.25) {};
		\node [style=none-small] (25) at (2, 4.5) {$\alpha_i'=\alpha_i+k\pi$};
		\node [style=none-small] (27) at (9.25, 4.5) {$\gamma_i'=\gamma_i+j\pi$};
		\node [style=none-small] (28) at (3, 3.5) {$\beta_i'=\beta_i+(j+k+1)\pi$};
	\end{pgfonlayer}
	\begin{pgfonlayer}{edgelayer}
		\draw [style=hadamard edge] (3) to (4);
		\draw [style=hadamard edge] (3) to (6);
		\draw [style=hadamard edge] (3) to (5);
		\draw [style=hadamard edge] (2) to (5);
		\draw [style=hadamard edge] (2) to (3);
		\draw [style=hadamard edge] (2) to (0);
		\draw [style=hadamard edge] (2) to (1);
		\draw [style=hadamard edge] (6) to (5);
		\draw [style=hadamard edge] (0) to (5);
		\draw [style=hadamard edge] (8) to (14);
		\draw [style=hadamard edge] (14) to (13);
		\draw [style=hadamard edge] (13) to (8);
		\draw [style=hadamard edge] (8) to (12);
		\draw [style=hadamard edge] (12) to (9);
		\draw [style=hadamard edge] (9) to (14);
		\draw [style=gray edge] (21.center) to (23.center);
		\draw [style=gray edge] (23.center) to (24.center);
		\draw [style=gray edge] (24.center) to (22.center);
		\draw [style=gray edge] (22.center) to (21.center);
	\end{pgfonlayer}
\end{tikzpicture}}
	\caption{(a) Transformation of a circuit for arbitrary single-qubit rotations into an equivalent graph-like diagram. 
		(b) Local complementation (lc) and (c) Pivoting (p) rule applied on exemplary graph-like ZX-diagrams.}
	\label{fig:gl-rules}
\end{figure*}
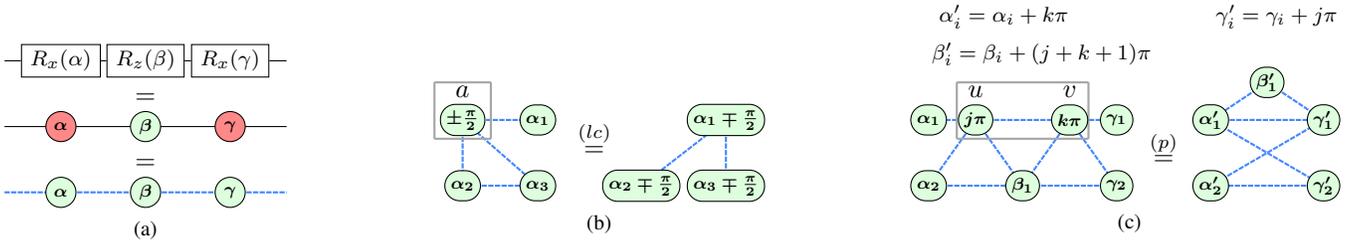

ZX-calculus is equipped with a set of sound and complete rules, allowing the transformation of ZX-diagrams into equivalent ones wrt. their linear map. 
Using those rules, we can transform any ZX-diagram into an equivalent \textit{graph-like} diagram, where we allow only Z spiders connected via Hadamard wires~\cite{duncan2020graph} (cf. \autoref{fig:gl-diagram}). 

These are especially useful for the one-way-model.
They can be directly interpreted as measurement patterns where the graph spanned by spiders and Hadamard wires corresponds to a graph state.
The spider's angles correspond to measurements in the equatorial basis of the Bloch sphere~\cite{Backens.2021}. 

Recent research has shown promising results in reducing the complexity of graph-like diagrams in terms of spiders and wires~\cite{duncan2020graph, staudacherreducing}.

This is mainly done by using two graph theoretic rewrite rules named Local complementation and Pivoting which we can use to eliminate most of the ``Clifford'' part in diagrams, i.e., spiders with an angle of $k \pi/2$, $k\in \mathbb{Z}$. 
Local complementation applied on a spider with an odd multiple of $\pi/2$ removes the spider and flips the Hadamard connections between the neighboring spiders, whereas Pivoting applied on a pair of spiders with an even multiple of $\pi/2$ removes the pair and flips the connections between different sets of neighboring spiders (cf. \autoref{fig:gl-rules-lc} and~\ref{fig:gl-rules-p}). 
Since every spider in a graph-like diagram corresponds to a qubit in the one-way model, we can use those rules to reduce the number of qubits needed for the computation. 

\section{Related Work}%
\label{sec:sota}
Several compiler frameworks and tools are currently available for executing algorithms on gate-based QPUs or annealing devices~\cite{fingerhuthopensourcequantum, LaRose2019, sivarajah2020t}.
Most high-level programming tools use sequential execution of basic quantum gates~\cite{chall+opp, oqasm, Qsharp-language, LaRose2019, fingerhuthopensourcequantum}.

The one-way quantum computing paradigm is fundamentally different from the standard quantum circuit model (see \autoref{sec:background}).
To allow integrating photonic one-way hardware into current programming frameworks and the HPC ecosystem, we require translation from the circuit model.

The framework by Zhang Hezi et al. \cite{PhotonicOneWayCompilerStBarb} aims particularly at mapping a quantum circuit to a specific photonic hardware architecture.
This hardware abstraction model relies on a 2D lattice arrangement of resource state generators (RSGs) that create a 3-qubit graph state in each clock cycle.

For translation, the scheme takes an input circuit and transpiles it to the universal gate set $\{J(\alpha), CZ\}$ where $(J\left(\alpha\right) \equiv H R_z(\alpha) )$ and construct an (in general) non-planar graph out of it.
They separate this graph into multiple planar sub-graphs, each of which may contain high-degree nodes, i.e., nodes with a large number of edges.
Further decomposition results in numerous appropriate low-degree (in their case, the 3-qubit GHZ-states) graph states and additional fusion operations.
A clock cycle marks the execution of one planar graph state mapped to the physical layer on the considered hardware.

\begin{figure*}[t!]
    \centering
	\input{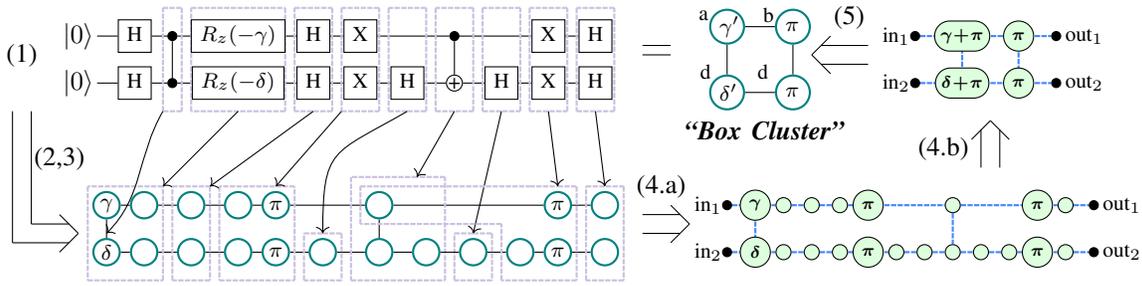}
	\caption{The ``Box'' Cluster for four physical qubits (centrally located) translates into full Grover search for two qubits. 
		Our pipeline takes the circuit description of this Grover algorithm as an input and translates it to the \textit{m-graph} which is then optimized and corresponds to a ``Box'' Cluster.
		The measurement angles $ \gamma',\delta' $ are adapted according to the hardware.
	}
	\label{fig:horse-shoe}\label{fig:grover}\label{fig:pipeline}
\end{figure*}

Vijayan Madhav Krishnan et al. \cite{RohdeGraphStateCompiler} propose a framework using the stabilizer formalism.
It converts a high-level description of a quantum circuit into a set of instructions for graph state preparation together with local operations and non-Pauli measurements. 
The approach is a modification of the Initialization-CNOT-Measurement (ICM) model, together with the single-qubit teleportation technique. 

First, they decompose the quantum circuit into Clifford and $T$ gate operations, the latter of which are performed via measurement-induced $T$ and $T^{\dagger}$ gate teleportation. 
The resulting intermediate circuit -- called \textit{inverse ICM decomposition} -- consists of a block of Clifford operations, Pauli corrections, and measurements in bases $\mathcal{B}_{j}(\pm \frac{\pi}{4})$. 
Since Clifford operations can be efficiently simulated, a classical computer calculates the output state of the Clifford block. The resulting stabilizer state translates to a graph state. 
With that at hand, the hardware instructions include the resulting graph state together with single-qubit unitaries (due to conversion) and non-Pauli measurements.


\section{Compiler}
\label{sec:method}

\subsection{Compiler pipeline}
To execute an arbitrary \textit{QASM} file on a photonic one-way processor, we perform the following translation and optimization steps (see also \autoref{fig:pipeline}):

\subsubsection*{\textbf{Step 1}} 
We transpile an arbitrary \textit{QASM} file as an input in the universal gate set $ \{R_x(\theta),R_z(\theta),H,CX,CZ\}$. 
Various tools for this were proposed in the past~\cite{Qsharp-language,compilerdesign-realhw,green2013quipper, qiskit}.

\subsubsection*{\textbf{Step 2}} 
We rewrite the circuit as a sequence of $HR_z(\theta)$ and $CZ$ operations as follows:

\begin{equation}
	\begin{aligned}
		R_z(\theta)&=[H R_z(0)][H R_z(\theta)]\\
		H&=H R_z(0)\\
		R_x(\theta)&=[H R_z(\theta)][H R_z(0)]\\
		CNOT&=[\mathbb{I}\otimes H] CZ [\mathbb{I}\otimes H]
	\end{aligned}
\end{equation}

\subsubsection*{\textbf{Step 3}} 
The \textit{m-graph} is defined by the sequence of $H R_z(\theta)$ and $CZ$: 
The number of qubits in the circuit defines the number of rows of the \textit{m-graph}.
A horizontal edge between two vertices in the \textit{m-graph} corresponds to a logical $H R_z(-\alpha)$ operation and a physical measurement in the basis $\mathcal{B}_{j}(\alpha)$ on the next physical qubit to the right.
Similarly, a $CZ$ corresponds to a vertical connection between two qubits~\cite{Greganti2021}.

\subsubsection*{\textbf{Step 4}} 
We interpret this \textit{m-graph} as a graph-like (4.a) diagram in ZX-calculus and optimize it (4.b).

\subsubsection*{\textbf{Step 5}}
The resulting \textit{m-graph} defines the required input graph state.
Each vertex shows the angle $ (\alpha) $ and its position in the graph defines its order of execution.

\subsection{Optimization}

We optimize the \textit{graph-like} diagram of ZX-Calculus with phase teleportation from \cite{kissinger2020reducing} to eliminate redundant non-Clifford spiders, followed by the algorithm from \cite{duncan2020graph} to eliminate Clifford spiders. 
Since both optimization strategies preserve the graph-like property, we can still interpret the diagram as \textit{m-graph}.

\section{Hardware Instructions}
\subsection{Cluster state generation}
For universal computation, we need an arbitrary connected graph state as an input.
In particular, the photonic setup in \autoref{fig:setup} generates $ \ket{\ghz} = \frac{1}{\sqrt{2}} (\ket{0000} + \ket{1111}) $ and \mbox{$ \ket{\psi} = \frac{1}{2} (\ket{0000} + \ket{0011} + \ket{1100} - \ket{1111})$} as input states~\cite{JWPanSetup}.
By activating%
\footnote{\label{fn:hwp}In an actual hardware setup, \textit{activating} corresponds to an angle of $ 22.5^{\circ} $, while \textit{deactivating} corresponds to $ 0^{\circ} $. For simulation, we can certainly just remove the HWP from the optical circuit.} %
or deactivating the half-wave plate (HWP) between the two bottom polarizing beam splitters (PBS), we produce the $ \ket{GHZ} $ or $ \ket{\psi} $ states, respectively.
With that, we can realize arbitrary 4-photon graph states of \autoref{fig:cluster-states} as such:

\begin{description}[labelwidth=1cm]
	\item[\textbf{Linear:}] The 4-qubit linear cluster is equivalent to $ \ket{\psi} $ up to a local operation $ H_1H_4 $ (see row 1, column 1 in \autoref{fig:cluster-states}). 
	Up to additional local (1-qubit + \textit{SWAP}) operations, it is also equivalent to arbitrary connected 4-qubit clusters (e.g., the ``Box'', see row 1 in \autoref{fig:cluster-states}).
	The only exceptions are the GHZ graph and the fully connected graph (see row 2 in \autoref{fig:cluster-states})
	\item[\textbf{GHZ:}] In the 4-qubit GHZ cluster, all nodes connect towards a single node (see row 2, column 1 in \autoref{fig:cluster-extraction}).
	Up to local (1-qubit) operations, it is equivalent to the $ \ket{\ghz} $ state and to a fully connected 4-qubit graph (see row 2 in \autoref{fig:cluster-states})
\end{description}

\noindent
In theory, the setup allows extension for producing $ \ket{\ghz} $ and $ \ket{\psi} $ states for an arbitrary number of qubits, however, for larger graphs, other setups are more suitable~\cite{Thomas.2022}.
Please note that for three qubits, the GHZ and linear cluster states are identical and GHZ with $ >4 $ qubits does not fit the definition of a cluster state.

\subsection{Graph state generation with fusion}
After the optimization in Step~4, we are left with a \textit{m-graph} that is not locally complementary to the GHZ or linear graph state.
Hence, we decompose the \textit{m-graph} to a set of GHZ- and linear graph states, respectively.
The extraction works as follows:
\begin{enumerate}
	\item Find all nodes that have $ >2 $ edges and order them with respect to number of their direct neighbors. Those are the nodes which will become ``roots'' of a GHZ graph state.
	\item Extract all GHZ graph states and remove them from the ZX-graph until no node with $ >2 $ neighbors is left.
	\item All other spiders that have $ \leq2 $ neighbors are considered to be part of linear graph states and are extracted until the graph is empty.
\end{enumerate}

\begin{figure}[t]
	\centering
	\begin{tikzpicture}
	\begin{pgfonlayer}{nodelayer}
		\node [style=Z dot] (0) at (0.5, 2.5) {};
		\node [style=Z phase dot] (1) at (0.5, 1.5) {$\frac{\pi}{4}$};
		\node [style=Z phase dot] (2) at (1.75, 0.75) {$\frac{\pi}{4}$};
		\node [style=Z dot] (3) at (3.75, 3.5) {};
		\node [style=Z phase dot] (4) at (3.75, 2.5) {$\frac{7\pi}{4}$};
		\node [style=Z dot] (5) at (3.25, 0.5) {};
		\node [style=Z phase dot] (6) at (4.5, 0.5) {$\frac{7\pi}{4}$};
		\node [style=Z dot] (7) at (4.5, 1.75) {};
		\node [style=Z phase dot] (8) at (2, 4) {$\frac{\pi}{4}$};
		\node [style=Z dot] (9) at (2, 3) {};
		\node [style=cluster small] (27) at (8.25, 1.75) {};
		\node [style=cluster] (29) at (9, 0.5) {$\frac{\pi}{4}$};
		\node [style=cluster small] (32) at (10.25, 0.5) {};
		\node [style=cluster small] (34) at (9.5, 2) {};
		\node [style=cluster small] (36) at (10.25, 1.75) {};
		\node [style=cluster small] (43) at (11.75, 0.5) {};
		\node [style=cluster] (44) at (13, 1) {$\frac{7\pi}{4}$};
		\node [style=cluster small] (45) at (11.5, 1.75) {};
		\node [style=cluster small] (47) at (13, 2) {};
		\node [style=cluster] (57) at (12.25, 4.25) {$\frac{\pi}{4}$};
		\node [style=cluster small] (58) at (12.25, 3.25) {};
		\node [style=none-small-gray] (59) at (2, 4.75) {8};
		\node [style=none-small-gray] (60) at (1.5, 3) {9};
		\node [style=none-small-gray] (61) at (0, 2.75) {0};
		\node [style=none-small-gray] (62) at (0, 1) {1};
		\node [style=none-small-gray] (63) at (1.75, 0) {2};
		\node [style=none-small-gray] (64) at (3.25, 0) {5};
		\node [style=none-small-gray] (65) at (5.25, 0.5) {6};
		\node [style=none-small-gray] (66) at (5, 1.75) {7};
		\node [style=none-small-gray] (67) at (4.5, 2.5) {4};
		\node [style=none-small-gray] (68) at (3.75, 4) {3};
		\node [style=none-small-gray] (69) at (8.25, 0.5) {2};
		\node [style=none-small-gray] (70) at (10.25, 0) {5};
		\node [style=none-small-gray] (71) at (10.25, 1.25) {7};
		\node [style=none-small-gray] (72) at (9.5, 2.5) {9};
		\node [style=none-small-gray] (73) at (7.75, 1.75) {0};
		\node [style=none-small-gray] (74) at (13, 0.25) {6};
		\node [style=none-small-gray] (75) at (13.5, 2) {9};
		\node [style=none-small-gray] (76) at (11.75, 0) {5};
		\node [style=none-small-gray] (77) at (11.5, 1.25) {7};
		\node [style=none-small-gray] (78) at (12.25, 5) {8};
		\node [style=none-small-gray] (79) at (12.75, 3.25) {9};
		\node [style=cluster small] (80) at (8.25, 3) {};
		\node [style=cluster] (81) at (8, 4) {$\frac{\pi}{4}$};
		\node [style=cluster small] (83) at (9.5, 4.5) {};
		\node [style=cluster] (84) at (9.75, 3.5) {$\frac{7\pi}{4}$};
		\node [style=none-small-gray] (92) at (7.75, 3) {0};
		\node [style=none-small-gray] (93) at (8, 4.75) {1};
		\node [style=none-small-gray] (98) at (10.5, 3.5) {4};
		\node [style=none-small-gray] (99) at (9.5, 5) {3};
		\node [style=none] (100) at (5.75, 1.5) {};
		\node [style=none] (101) at (5.75, 2) {};
		\node [style=none] (102) at (6.75, 1.5) {};
		\node [style=none] (103) at (6.75, 2) {};
		\node [style=none] (104) at (7.25, 1.75) {};
		\node [style=none] (105) at (6.5, 2.25) {};
		\node [style=none] (106) at (6.5, 1.25) {};
	\end{pgfonlayer}
	\begin{pgfonlayer}{edgelayer}
		\draw [style=hadamard edge] (0) to (2);
		\draw [style=hadamard edge] (0) to (1);
		\draw [style=hadamard edge] (0) to (4);
		\draw [style=hadamard edge] (1) to (3);
		\draw [style=hadamard edge] (2) to (5);
		\draw [style=hadamard edge] (2) to (7);
		\draw [style=hadamard edge] (2) to (9);
		\draw [style=hadamard edge] (3) to (4);
		\draw [style=hadamard edge] (5) to (6);
		\draw [style=hadamard edge] (6) to (7);
		\draw [style=hadamard edge] (6) to (9);
		\draw [style=hadamard edge] (8) to (9);
		\draw (27) to (29);
		\draw (29) to (32);
		\draw (29) to (34);
		\draw (29) to (36);
		\draw (43) to (44);
		\draw (44) to (45);
		\draw (44) to (47);
		\draw (57) to (58);
		\draw (80) to (81);
		\draw (80) to (84);
		\draw (81) to (83);
		\draw (83) to (84);
		\draw [style=boldedge red] (80) to (27);
		\draw [style=boldedge red] (34) to (58);
		\draw [style=boldedge red] (58) to (47);
		\draw [style=boldedge red] (32) to (43);
		\draw [style=boldedge red] (36) to (45);
		\draw [style=boldedge] (100.center) to (102.center);
		\draw [style=boldedge] (101.center) to (103.center);
		\draw [style=boldedge] (106.center) to (104.center);
		\draw [style=boldedge] (104.center) to (105.center);
	\end{pgfonlayer}
\end{tikzpicture}
	\caption{Decomposition of an arbitrarily connected ZX-graph to a set of graph states, the respective measurement angles, and fusion operations (red connections).
		This case shows two linear (nodes 0-1-3-4 and 8-9) and two GHZ graph states (nodes 2-0-5-7-9 and 6-5-7-9). 
		To generate the actual graph state, the nodes that occur multiple times in different graphs (nodes 0, 5, 7, 9) are fused.
	}
	\label{fig:cluster-extraction}
\end{figure}
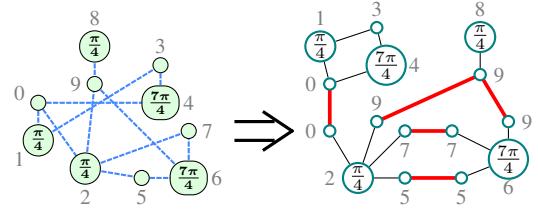

\autoref{fig:cluster-extraction} shows the decomposition of a ZX-graph that was extracted from the ``bell-n4'' circuit from \textit{QASMBench}~\cite{li2021qasmbench} with four logical qubits and 22 gates (of which 7 are two-qubit gates).
After optimization, we are left with a ZX-graph of 10 spiders, which is decomposed into 4 graph states that can be implemented on hardware such as the one in \autoref{fig:setup}.

Intersecting nodes between two graph states are connected using fusion gates~\cite{Browne.2005, Bartolucci.2021}.
\autoref{fig:cluster-extraction} shows that the photons labeled with ``0'', ``5'', ``7'' and ``9'' occur in multiple graph states.
These labels identify the fusion gates that must be performed on hardware after the GHZ and linear graph states are generated.
To perform fusion, one (in case of Type-1) or two (in case of Type-2) photons are measured, and we are left with one photon that connects two clusters.
That way, we can construct arbitrary input states with any hardware setup that is able to produce GHZ and linear graph states.

\subsection{Measurements basis}
The measurement basis of each photon (1-4) on the right of \autoref{fig:setup} is defined by the angle of the corresponding vertex in the \textit{m-graph}.
As an example, node ``a'' in the ``Box Cluster'' in \autoref{fig:pipeline} configures the angles of the QWP/HWP~``a'' in \autoref{fig:setup}.


\begin{table*}[t]
	\caption{Results of compiled measurement-patterns for several circuits of the OpenQASM benchmark. 
		The columns are structured as follows: \textbf{QASM-circuit:} name of the circuit in the OpenQASM benchmark, 
		\textbf{GHZ:} Number of GHZ graphs, \textbf{Linear:} Number of linear graphs, \textbf{Photons:} Number of total photons, \textbf{Comp.:} Number of optical components. 
		We compare with the circuital-model translation as built-in function of \textit{Perceval}}
	\label{tab:evaluation}
	\centering
	\begin{tabular}{l||c|c|c|c||c|c}
	                                                       &                \multicolumn{4}{c||}{\textbf{Ours}}                 & \multicolumn{2}{c}{\textbf{Perceval}} \\
	\rule{0pt}{1\normalbaselineskip} \textbf{QASM-circuit} & \textbf{GHZ} & \textbf{Linear} & \textbf{Photons} & \textbf{Comp.} & \textbf{Photons} &    \textbf{Comp.}    \\ \hline
	adder-n4                                               &      6       &        4        &        37        &      167       &        46        &         182          \\
	bell-n4                                                &      2       &        2        &        18        &       76       &        34        &         135          \\
	cat-state-n4                                           &      0       &        1        &        6         &       17       &        18        &          42          \\
	deutsch-n2                                             &      0       &        1        &        3         &       8        &        6         &          11          \\
	fredkin-n3                                             &      4       &        3        &        29        &      126       &        36        &         144          \\
	grover-n2                                              &      0       &        1        &        4         &       11       &        10        &          35          \\
	hs4-n4                                                 &      1       &        2        &        9         &       31       &        22        &          74          \\
	iswap-n2                                               &      0       &        1        &        3         &       8        &        10        &          32          \\
	linearsolver-n3                                        &      2       &        1        &        11        &       44       &        20        &          72          \\
	qft-n4                                                 &      3       &        3        &        34        &      147       &        54        &         236          \\
	teleportation-n3                                       &      0       &        1        &        4         &       11       &        12        &          29          \\
	toffoli-n3                                             &      2       &        2        &        23        &       91       &        28        &         112          \\
	wstate-n3                                              &      3       &        3        &        29        &      122       &        40        &         176          \\ \hline\hline
	adder-n10                                              &      25      &       10        &       218        &      1210      &        --        &          --          \\
	bv-n14                                                 &      1       &        0        &        15        &       43       &        54        &         119          \\
	multiply-n13                                           &      17      &        9        &       157        &      754       &       106        &         380          \\
	multiplier-n15                                         &      51      &       19        &       767        &      4706      &       474        &         2096         \\
	qf21-n15                                               &      18      &       14        &       299        &      1467      &       260        &         1148         \\
	qft-n15                                                &      14      &       14        &       462        &      2234      &       450        &         2097         \\
	qpe-n9                                                 &      11      &        7        &       121        &      568       &       104        &         432          \\
	sat-n11                                                &      68      &       34        &       834        &      4836      &        --        &          --          \\
	simon-n6                                               &      2       &        2        &        15        &       65       &        40        &         140 
\end{tabular}

\end{table*}

\section{Implementation and Evaluation}
We implement the presented pipeline with the following libraries:
\begin{itemize}
	\item \textbf{Qiskit}~\cite{qiskit} to transform an arbitrary circuit to the desired gate set (Step 1),
	\item \textbf{PyZX}~\cite{kissinger2020Pyzx} to translate the circuit to a graph-like diagram and optimize (Step 2-4), and 
	\item \textbf{Perceval}~\cite{perceval_2022} to create a photonic circuit as in \autoref{fig:setup}, that simulates the graph states and fusion operations.
\end{itemize} 
Our software is accessible on GitHub\footnote{\url{https://github.com/CDL-Uni-Vienna/photonq-compiler}} and includes a simulation of the execution of the hardware instructions with \textit{Perceval}.
We also include a parser from \textit{QASM} files into the basic gate set $ \{H R_z(\alpha), CZ \} $.\\

We show the differences between our resulting optical circuits and those provided by the converter package from \textit{Perceval}.
The latter translates \textit{Qiskit} quantum circuits into standard optical circuits with probabilistic photonic gates.
However, it should be noted that this evaluation has only limited expressiveness, as the one-way model and circuital model are conceptually very different.
The strength of the one-way model is not necessarily the reduction of the number of photons, but the fact that the overall computation can be performed deterministically.

\subsection{Simulation}
We simulate the photonic architecture depicted in \autoref{fig:setup} with \textit{Perceval} using standard optical components and extend the setup for $ >4 $ photons.
To connect two graph states as in \autoref{fig:cluster-extraction}, we implement fusion gates using single polarizing beam splitters and half-wave plates.
We choose to work with \mbox{Type-1} fusion gates, for which one of the fused photons is measured; the fusion is successful if the detector registers a single photon.

\subsection{Evaluation}
We apply our compiler to extract \textit{m-graphs} from circuits given by the \textit{QASMBench}~\cite{li2021qasmbench} project to demonstrate the validity of our pipeline and implementation. We execute these \textit{m-graphs} on a simulation of our photonic setups, such as that shown in \autoref{fig:setup} in the case of four photons.
\autoref{tab:evaluation} shows the number of required GHZ and linear graph states, together with the number of photons and optical elements -- the latter are counted with \textit{Perceval}.

First and foremost, it shows that we were able to decompose all circuits from this project and translate them to a photonic setup -- which is not true for the converter provided by \textit{Perceval}~\cite{convertorsperceval}.

Second, we compare the number of required photons and optical elements against the quantum circuit conversion provided by \textit{Perceval}.
For some cases  (e.g. \textit{grover-n2}, \textit{hs4-n4} or \textit{bv-n14}), \autoref{tab:evaluation} shows the full strength of the one-way model, where a single input graph state can perform rather complex calculations.
However, especially larger circuits with complex calculation show the excessive use of photons in the one-way model.

\section{Conclusion}%
\label{sec:conclusion}
Photons are well-known candidates for quantum computing, but they are currently underrepresented in publicly available environments, such as HPC centers.
The well-established quantum circuit model, however, is impracticable for photonic hardware.
Hence, measurement-based quantum computing (MBQC) is the only feasible alternative for large-scale photonic quantum computing.
The quantum circuit model and the measurement-based one-way model of quantum computation are known to be computationally equivalent.
Still, compilers that translate circuits into the one-way model are not established yet~\cite{PhotonicOneWayCompilerStBarb, RohdeGraphStateCompiler}.

In this paper, we describe a pipeline for a compiler that converts an arbitrary quantum circuit to an optimized instruction set for a one-way QPU. 
We demonstrate the translation to a discrete variable photonic platform, but the concept should be applicable to other quantum computing technologies that implement the one-way model, e.g., superconducting qubits or trapped ion qubits~\cite{RealizationsMBQC, MBQCIons}.

Our implementation constructs a \textit{measurement graph} from the \textit{QASM} input by using simple transformation rules and optimizes it with the built-in methods of ZX-Calculus.
We find that ZX-Calculus is ideally suited to extract specific hardware instructions from \textit{graph-like} diagrams.

This work is only a first demonstration of the power of a ZX-Calculus-based compiler for photonic QPUs.
We plan to further investigate the optimization for specific hardware architectures and other platforms, such as the recent proposal where 14-photon GHZ and 12-photon linear states were experimentally realized~\cite{Thomas.2022}.

\section*{Acknowledgment}
The authors would like to thank Francesco Giorgino for the fruitful discussions and the patient help with some of the theory.

This work is partially funded from the Austrian Federal Ministry for Digital and Economic Affairs, the National Foundation for Research, Technology and Development and the Christian Doppler Research Association.
This work is additionally partially supported by the German Federal Ministry of Education and Research (BMBF) under the funding program Quantum Technologies - From Basic Research to Market under contract number 13N16077.


\end{document}